\documentclass[pra,aps,twocolumn,amsmath,amssymb,superscriptaddress,floatfix]{revtex4-2}
\usepackage{amsthm}
\usepackage{graphicx}
\usepackage{color}
\usepackage{comment}

\begin{document}

\title{Dispersion engineering in spin-orbit coupled spinor $F=1$ condensates driven by negative masses}

\author{K. Rajaswathi}
\affiliation{Department of Physics, Bharathidasan University, Tiruchirappalli 620024, Tamil Nadu, India}

\author{S. Bhuvaneswari}
\affiliation{Centre for Nonlinear Science (CeNSc), Government College for Women, Kumbakonam 612001, Tamil Nadu, India}

\author{R. Radha}
\affiliation{Centre for Nonlinear Science (CeNSc), Government College for Women(Autonomous), Kumbakonam 612001, Tamil Nadu, India}

\author{P. Muruganandam}
\affiliation{Department of Physics, Bharathidasan University, Tiruchirappalli 620024, Tamil Nadu, India}

\date{\today}

\begin{abstract}
In this paper, we bring out several potential signatures of negative mass regimes while investigating an expanding spin-orbit (SO) coupled spinor $F=1$ Bose-Einstein condensates by analyzing the dispersion relation of the single-particle quantum system. In SO-coupled spinor condensates, a negative mass parameter generates a wave packet that propagates in the opposite direction of the momentum. We analyze the dynamics of spin waves analytically and present a simple approach to investigate the expansion of spinor condensates. In particular, we examine the dynamics when both masses  are negative, which results in the spinor condensates splitting into two counter-propagating self-interfering packets (SIPs). Using numerical simulations of the coupled Gross-Pitaevskii equations, we demonstrate the density expansion and self-interference patterns with and without magnetization for repulsive and attractive interactions  with different coupling parameters. The highlight of  our investigation is that we are able to unearth several phenomena observed in experiments, such as self-interfering packets, pileup, modulation instability, slow down, self-trapping, and gap solitons. In particular, the gap soliton exists  at the gap created by the intersection of two negative masses. 
\end{abstract}  

\maketitle

\section{Introduction}

The interaction between a quantum particle's spin and momentum, known as spin-orbit coupling (SOC), has opened up numerous avenues in quantum science~\cite{Juzeliunas2006, Lin2009}. This peculiar yet fascinating field of research finds applications in almost all branches of physics like  for example, topological insulators~\cite{Hasan2010, Qi2011}, spin Hall effect~\cite{Klitzing1986}, Majorana fermions~\cite{Wilczek2009}, spintronic devices~\cite{Koralek2009}, and quantum computing applications~\cite{Alicea2011}. Recently, Bose-Einstein condensation (BEC), which belongs to the class of ultracold atoms, has provided an excellent platform for conducting quantum studies beyond natural conditions. One of the major challenges in this investigation is that the atoms are neutral, and coupling with the gauge field requires engineering synthetic gauge fields. To overcome this inherent issue of neutrality, several proposals have been put forward to realize synthetic gauge fields for quantum gases~\cite{Lin2009a, Paredes2004, Kinoshita2004, Hadzibabic2006}.

In the past few years, research on synthetic gauge fields has evolved towards the \textit{``on-demand''} engineering of spin-orbit coupling (SOC) using laser beams. This field of research is promising due to its ability to realize exotic configurations of nontrivial topology and simulate vital electronic phenomena in condensed matter physics. The Spielman group at NIST made a seminal contribution to this exciting field by engineering the SOC in neutral Bose-Einstein condensates (BECs) using a pair of lasers to dress two atomic spin states~\cite{Lin2011, Lin2009}. They achieved momentum-sensitive coupling in $^{87}$Rb, which has equal contributions from Rashba and Dresselhaus, by using a pair of Raman lasers to address two of its $F=1$ hyperfine spin states: $\lvert \uparrow \rangle = \lvert F=1, m_{F}=0 \rangle$ and $\lvert \downarrow \rangle = \lvert F=1, m_{F}=-1 \rangle$.

Following the pioneering work of the Spielman group at NIST~\cite{Lin2011}, synthetic SOC has been successfully engineered with both neutral bosonic and fermionic ultracold atoms~\cite{Gong2011}, which not only exhibits many exotic phases, but also opens up a lot of avenues to explore novel SOC physics with an unprecedented level of tunability of experimental parameters. Taking advantage of the exceptional tunability of experimental parameters, SOC in BECs presents an exciting possibility of engineering more complex dispersion relations through controlling the Raman laser setup. One of the fundamental studies in this direction is the concept of negative mass by suitably engineering the dispersion relation in diverse quantum systems. The negative mass is a hypothetical concept of matter whose mass is of opposite sign to the mass of normal matter, say, for example, $-2$kg. Such matter may violate some energy conditions and exhibit strange characteristics~\cite{Eiermann2003}. Recently, Khamehchi \emph{et al.} have shown how the peculiar dispersion relation of an atomic spin-orbit coupled Bose-Einstein condensate could lead to unconventional wave packet dynamics, interpreted as negative-mass hydrodynamics and reported phenomena such as self-trapping, soliton trains, and dynamical instabilities~\cite{Khamehchi2017}. Zhao \emph{et al.}~\cite{Zhao2020} reported a study on the periodic transition between negative and positive inertial mass with AC oscillation of a spin soliton driven by a constant force. They further discussed the weak force that could be diagnosed from the AC oscillation phenomena of the spin soliton, which is similar to recent experiments observed in optomechanical instruments. Farolfi \emph{et al.} investigated the collisional dynamics of magnetic solitons in a harmonically trapped binary mixture using phase imprinting~\cite{Farolfi2020}. Following their work, Chai \emph{et al.} reported a magnetic soliton in  a spin-1 Bose-Einstein condensate  using the magnetic phase imprinting method and observed good agreement with  numerical simulations based on the one-dimensional Gross-Pitaevskii equation~\cite{Chai2020}. Meng \emph{et al.} investigated spin solitons employing the dispersion relation with critical velocities to demarcate the boundary between negative and positive mass regimes. A correlation between  the width and speed helped them to distinguish between bright and dark solitons ~\cite{Meng2022}.

A more comprehensive theoretical study on the negative mass effect in SO-coupled BECs was reported recently by Colas \emph{et al.}~\cite{Colas2018}. In this work, the authors have discussed the conceptual evidence and the physical interpretation of the negative masses beyond the description reported in Ref.~\cite{Khamehchi2017}. They have also brought out  the characteristics of several regimes that determine the signs of different effective mass parameters, say $m_1$ and $m_2$ derived from energy-dispersion relation. At this juncture, it is worth pointing out that both these investigations on the negative mass effect were centered around SO-coupled spin-$1/2$ BECs, which means that the ramifications of negative mass on SOC spinor $F=1$ BECs have still not yet been explored. Thus, inspired by the distinct features of SO coupling and the physical relevance of the $F=1$ system, we intend to study the dynamical behavior of negative mass in SO-coupled BECs. In this paper, we  investigate the impact of negative mass in spin-orbit coupled $F=1$ spinor condensates. First, we discuss the situation at the single-particle level showing the peculiar features of the single-particle energy spectrum. We then study the dynamics in a regime where both the masses, $m_1$ and $m_2$, are negative. 

The paper is organized as follows: After a detailed introduction, Sec.~\ref{sec2} presents the theoretical model that describes the concept of negative mass and the Hamiltonian of the problem under investigation. The energy dispersion relation and the characteristics of the dispersion curve are discussed in Sec.~\ref{sec3}. In Sec.~\ref{sec4}, we bring out through numerical simulation several interesting signatures of negative mass regimes like symmetric expansion, self-interference pattern (SIP), etc. We then conclude with the highlights of the investigation in Sec.~\ref{sec5}.

\section{Theoretical Model}
\label{sec2}
We review the fundamental concepts of negative mass using the energy dispersion relation~\cite{Pethick2008}. Expanding the energy dispersion relation up to the second-order, we get $E(p) = E_0 + v_g (p-p_0) + (p-p_0)^2 / [ 2 m_2 (p_0)]$, which can be used to deduce two mass parameters $m_1$ and $m_2$ that dictate the dynamics of the system under consideration by %
\begin{subequations} 
\label{massparameters}
\begin{align}
m_1 &= \frac{p}{v_g} = p \left( \frac{\partial E}{\partial p} \right)^{-1}
\label{m1_label} \\
m_2 & = \left( \frac{\partial^2 E}{\partial p^2} \right)^{-1}.
\label{m2_label}
\end{align}
\end{subequations}%
Both mass parameters are equally important if one has to consider both the propagation and diffusion of wave packets. In an isotropic system, the mass parameters $m_1$ and $m_2$ are respectively related to the group velocity and the acceleration of wave packets as given by Eq.~\eqref{massparameters}. In the case of anisotropic materials, the above relations hold with a slight modification such that the particle momentum and velocity can be related as $p_i=m_{ij}v_j$, where $m_{ij}$ is the effective mass tensor, $p_i$ and $v_j (i, j = x, y, z)$ are components of momentum and velocity, respectively. For the present study, we adopt the experimental realization of tunable SO-coupled BEC in $^{87}$Rb reported by Lin \emph{et al.}~\cite{Lin2011}. In that configuration, two counter-propagating Raman lasers of wavelength ($\lambda_r$) were used to couple the states with strength $\Omega$. The Raman wave vector is given by $k_L = 2 \pi \sin(\beta_r/2)/\lambda_r$, where $\beta_r$ is the orientation of Raman lasers. %

In this context, we consider the SO coupling among the three spin components of the $F=1$ hyperfine state 5S$_{1/2}$ of $^{87}$Rb, namely, $\lvert F=1, m_{F}=1 \rangle$, $\lvert F=1, m_{F}=0 \rangle$ and $\lvert F=1, m_{F}=-1 \rangle$, where $m_F$ is the $z$ projection of $F$ \cite{Gautam2014}. The observation of Feshbach resonances in $^{87}$Rb to manipulate scattering lengths has been extensively analyzed earlier \cite{Newbury1995, Vogels1997, Compton2012}. Then, the single-particle Hamiltonian of the quasi hyperfine spin-1 SO-coupled BEC confined along the $x$ axis by a strong transverse trap along $y$ and $z$ axes can be written as
\begin{align}
H_0=\frac{p_{x}^{2}}{2m}+k_{L} p_{x}\varSigma _{z}+V(x) +\Omega \varSigma _{x}. \label{eq:hamil}
\end{align}%
We consider three possible SO couplings in the above Hamiltonian of the form  $k_{L} p_{x}\varSigma _{x}$, $k_{L} p_{x}\varSigma _{y}$ and $k_{L} p_{x}\varSigma _{z}$, where $k_{L}$ is the SO coupling parameter, $\Omega$ is the Rabi frequency, $V(x)$ is the trapping potential, $p_x=-\mathrm{i}\hbar\partial_{x}$ is momentum operator, $\varSigma_{x}$, $\varSigma_{y}$ and $\varSigma_{z}$ are the spinor-1 angular momentum operators which are given by %
\begin{align} &
\varSigma _{x} =\frac{1}{\sqrt{2}}
\begin{pmatrix}
0 & 1 & 0 \\
1 & 0 & 1 \\
0 & 1 & 0
\end{pmatrix}, \;\;
\varSigma _{y}=\frac{\mathrm{i}}{\sqrt{2}}
\begin{pmatrix}
0 & -1 & 0 \\
1 & 0 & -1 \\
0 & 1 & 0
\end{pmatrix}, 
\notag \\
& \varSigma _{z}  =
\begin{pmatrix}
1 & 0 & 0 \\
0 & 0 & 0 \\
0 & 0 & -1
\end{pmatrix}.
\end{align}
In the standard computational basis, the matrix notation of the Hamiltonian is
\begin{align}
H_{0}=
\begin{pmatrix}
\displaystyle\frac{p_x^{2}}{2m}+p_x k_{L}  & \displaystyle\frac{\Omega }{\sqrt{2}} & 0 \\
\displaystyle\frac{\Omega }{\sqrt{2}} & \displaystyle\frac{p_x^{2}}{2m} & \displaystyle\frac{\Omega }{\sqrt{2}} \\
0 & \displaystyle\frac{\Omega }{\sqrt{2}} & \displaystyle\frac{p_x^{2}}{2m}-p_x k_{L}
\end{pmatrix}.
\label{Hamiltonian}
\end{align}
By setting the trapping potential zero, energy eigenspectrum corresponding to the homogeneous non-interacting SO coupled BEC can be written as
\begin{align}
\omega_{y, z}(p)=\frac{p_x^{2}}{2m} \;\; \text{and} \; \omega_{\pm y, z}(p) =\frac{p_x^{2}}{2m}\pm \sqrt{\Omega^{2}+p_x^{2}k_{L}^{2}}. \label{energy-dis}
\end{align} %
If we consider SO coupling along the $y$ or $z$ axis, the eigenvalues remain the same. On the other hand, the energy of the system considering SO coupling along $x$ axis is given by 
\begin{align}
\omega_{x}(p)=\frac{p_x^{2}}{2m} \; \text{and} \; \omega_{\pm x}(p) =\frac{p_x^2}{2 m}\pm k_{L} p_x-\Omega. \label{energyxx-dis}
\end{align}
Under the Hartree approximation, the spinor BEC in a quasi-1D trap can be described by a set of three coupled Gross-Pitaevskii equations for the three components of wave function $\psi_{j}$, $j = -1, 0, +1$ as \cite{Ho1998, Salasnich2002, Kawaguchi2012, Gautam2015, Ravisankar2021} %
\begin{subequations}\label{eq:gpe}
\begin{align}
\mathrm{i} \hbar \frac{\partial \Psi_{\pm 1}}{\partial t} = & \bigg[ -\frac{\hbar^2}{2m}\frac{\partial^2}{\partial x^2} + V(x) + c_{0} n + c_{2} (n_{\pm 1} + n_0 \notag \\ & -n_{\mp 1}  \bigg] \Psi_{\pm 1}  
+ c_{2} \Psi_0^2 \Psi_{\mp 1}^* + \frac{\Omega}{\sqrt{2}}\Psi_0 \mp \mathrm{i}\hbar k_{L} \frac{\partial\Psi_{\pm1}}{\partial x}, \\ 
\mathrm{i} \hbar\frac{\partial \Psi_0}{\partial t} = & \left[ -\frac{\hbar^2}{2m}\frac{\partial^2}{\partial x^2} + V(x) + c_{0} n  + c_{2} (n_{+1}+n_{-1}) \right]\Psi_0 \notag\\&
+ 2c_{2} \psi_{+1}\Psi_{-1}\Psi_0^* + \frac{\Omega}{\sqrt{2}}(\Psi_{+1}+\Psi_{-1}), 
\end{align}
\end{subequations}%
with
\begin{align}
c_0 = \frac{2\hbar^2(a_{0}+2a_{2})}{3ml_{yz}^2 }, \; 
c_2 = \frac{2\hbar^2(a_{2}-a_{0})}{3ml_{yz}^2},
\end{align}
where $a_0$ and $a_2$ are the $s$-wave scattering lengths in the total spin $0$ and $2$ channels, respectively. $n =\sum_{j} n_j$ is the total density with $n_j = \vert \Psi_j\vert ^2$, $j=1, 0, -1$ are the densities of the individual spin components.

We choose spin-1 $^{87}$Rb atoms  with scattering lengths $a_{0} = 101.8a_{B}$ and $a_{2} = 100.4a_{B}$, where $a_{B}$ is the Bohr radius  and use the experimental trapping frequency range  2$\pi$ $\times$ 230 Hz~\cite{Anker2005} in our numerical simulations. If $c_{0} > 0$ and $c_{2} < 0$, the interaction is repulsive, while it is attractive for $c_{0} < 0$ and $c_{2} < 0$~\cite{Ravisankar2021, Pardeep2021} and the concept of Feshbach resonance can be employed for manipulating the scattering lengths ~\cite{Inouye1998, Marte2002} to determine the characteristics of the condensates~\cite{Chin2010}.

The harmonic trap is given as, $V(x) = m \omega_x^2 x^2/2$ and $l_{yz} = \sqrt{\hbar/(m\omega_{yz})}$ is the oscillator length in the transverse $y-z$ plane, where $\omega_{yz} = \sqrt{\omega_y\omega_z}$. The normalization condition is 
\begin{align}
\int_{-\infty}^{\infty}\,dx\,\sum_{j=-1}^{1}\,\vert\Psi_{j}(x)\vert^2 = N. \label{eq:norm:1}
\end{align}
In the above, $N$ is the total number of atoms, which is of the order of $10^{3}$ and $l_{0} = \sqrt{\hbar/m \omega_{x}}$ is the oscillator length along the $x$-axis.

It is convenient to transform Eq.~\eqref{eq:gpe} into dimensionless form, for which we use the following change of variables, 
\begin{align}
\label{eq:Dvar}
x = l_{0} \tilde{x}, \;  t = \omega_{x}^{-1} \tilde{t}, \;  \Psi(x, t) = N^{\frac{1}{2}} l_{0}^{-\frac{1}{2}}
\psi_{j} (\tilde{x},\tilde{t}).
\end{align}
By applying the change of variables \eqref{eq:Dvar} to \eqref{eq:gpe}, the coupled GP equations can be expressed as,
\begin{subequations}\label{eq:gpe1}
\begin{align}
\mathrm{i} \frac{\partial \psi_{\pm 1}}{\partial t} =  & \left[ - \frac{1}{2}\frac{\partial^2}{\partial \tilde x^2} + \tilde{V} + \tilde c_{0} \tilde{n} + \tilde c_{2}  (\tilde{n}_{\pm 1} + \tilde{n_0}  -\tilde{n}_{\mp 1} ) \right] \psi_{\pm 1}  \notag \\ &
+ \tilde c_{2} \psi_0^2 \psi_{\mp 1}^* + \frac{\tilde{\Omega}}{\sqrt{2}}\psi_0 \mp \mathrm{i} \tilde{k_{L}} \frac{\partial\psi_{\pm1}}{\partial x}, \\
\mathrm{i}  \frac{\partial \psi_0}{\partial t} = & \left[ - \frac{1}{2} \frac{\partial^2}{\partial \tilde x^2} + \tilde{V}+\tilde c_{0} \tilde{n} +  \tilde c_{2} (\tilde{n}_{+1}+\tilde{n}_{-1}) \right]\psi_0 \notag\\&
+ 2\tilde c_{2} \psi_{+1}\psi_{-1}\psi_0^* + \frac{\tilde{\Omega}}{\sqrt{2}}(\psi_{+1}+\psi_{-1}), 
\end{align}
\end{subequations}%
with
\begin{align}
\tilde{c_0} = \frac{2N l_{0} (a_{0}+2a_{2})}{3l_{yz}^2 }, \; 
\tilde{c_2} = \frac{2N l_{0} (a_{2}-a_{0})}{3l_{yz}^2},
\end{align}
where $\tilde{V} = \tilde{x}^2/2$, $\tilde{k}_{L} = \hbar k_{r}/m \omega_{x}l_{0}$, $\tilde{\Omega} = \Omega / \hbar\omega_{x}$, $\tilde{n}_j = \vert \psi_j\vert^2$ with $j=1, 0, -1$ and $\tilde{n} = \sum_{j=-1}^{1} \vert \psi_j \vert^2$. 
Since the number of atoms $N$ is absorbed in the 
dimensionless quantities,
the normalization and magnetization conditions satisfied by the $\psi_{j}$'s become
\begin{align}
& \int_{-\infty}^{\infty}\,\sum_{j=-1}^{1}\,\tilde{n}_{j}(\tilde{x}){d\tilde{x}} = 1, \label{eq:norm:2} 
\end{align}%
and
\begin{align}
& M = \int_{-\infty}^{\infty}\,[{\tilde{n}_{1}({\tilde{x}})}-{\tilde{n}_{-1}({\tilde{x}})}]\,{d{\tilde{x}}}.
\end{align}%
For the sake of simplicity of notation, we denote the dimensionless variables without a tilde in the rest of the paper.
The numerical results are discussed in Sec.~\ref{sec4} for repulsive and attractive interactions with and without magnetization.

\section{Analytical Results}
\label{sec3}
The energy dispersion relation is the fundamental framework of the investigation which governs the dynamics of the underlying dynamical system. It is evident from Eq.~\eqref{energy-dis} that the energy spectrum consists of two branches. The $\pm$ sign denotes the different helicity basis corresponding to either parallel or anti-parallel spin-index with reference to the wave vector. It is interesting to note that in the conventional BECs, the atoms condense at the ground state often recognized as a non-degenerate zero-momentum state, which does not occur in the SO-coupled BECs where one comes across multiple lowest degenerate energy states due to nonparabolic energy-momentum dispersion,  unlike the conventional BECs. %
\begin{figure}[!ht]
    \centering\includegraphics[width=0.99\linewidth]{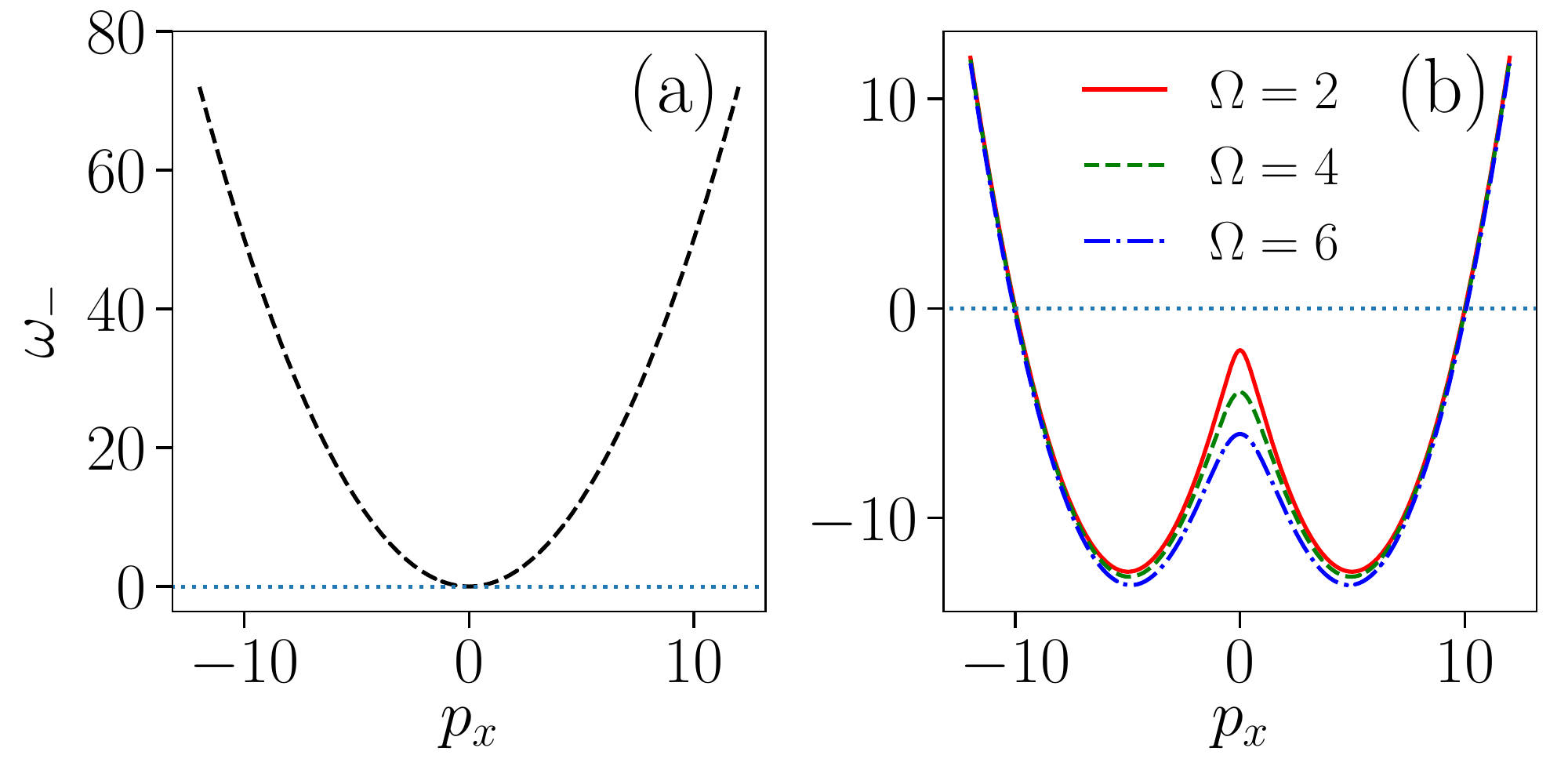}
    \caption{Plots of (a) energy dispersion in the absence of spin-orbit coupling and (b) energy dispersion as a function of momentum for some representative values of Rabi coupling with fixed parameters of $m=1$ and $k_{L}=5$.}
\label{Parabolic}
\end{figure}%
Figs.~\ref{Parabolic}(a) and \ref{Parabolic}(b) show the energy spectrum corresponding to BECs without  and with SO-coupling and Rabi coupling respectively. It is quite obvious from the above  that the energy spectrum exhibits a single free-particle parabolic dispersion demonstrating the lowest energy state in the absence of the spin-orbit and Rabi couplings, i.e. $\omega_{\pm} = p_x^2/2m$. In other words, without SO coupling, the Hamiltonian represented by Eq.~\eqref{Hamiltonian} has a unique minimum at $p_x = 0$ as shown in Fig.~\ref{Parabolic}(a). It is straightforward to note from the expressions of the mass parameters that purely parabolic dispersion amounts to equal values of $m_1$ and $m_2$. As evident from Fig.~\ref{Parabolic}(b), the SO coupling changes the energy spectrum considerably leading to a non-parabolic energy dispersion, where the two mass parameters cannot be equal. In the present case, we identify the non-parabolic energy dispersion as a consequence of the negative effective mass parameter which is related to the negative curvature of the dispersion relation. Another notable feature is that the introduction of Rabi coupling when reinforced with the SO coupling generates a symmetric double-well potential as shown in Fig.~\ref{Parabolic}(b) unlike the results reported in \cite{Colas2018}. The symmetric nature of the double-well potential can be attributed to the absence of detuning parameter in the Hamiltonian unlike in \cite{Colas2018}. It should also be reiterated that the SO coupling term has been added as a linear perturbation in the Hamiltonian given by Eq. (\ref{eq:hamil})  which contributes to  the double-well potential unlike in \cite{Su2021} where the SO coupling has been reinforced with momentum giving rise to a quadratic term in the Hamiltonian thereby generating a triple well structure. Dispersion with negative curvature plays an increasingly important role in quantum hydrodynamics, fluid dynamics, and optics. Manifestations of negative mass effects are observed in several quantum systems and the SOC-BECs can be exploited to witness controllable dispersion engineering through negative masses. In accordance with our results, more recently, a similar form of the double-well dispersion relation is identified \cite{Wang2023}. In contrast to the above, the standard spin $F=1$ BEC may have a three-well structure also. This arises due to the fact that the SO coupling term has been reinforced with momentum in the Hamiltonian, making it quadratic, which contributes to the three-well structure.

\begin{figure}[!ht]
\centering\includegraphics[width=0.99\linewidth]{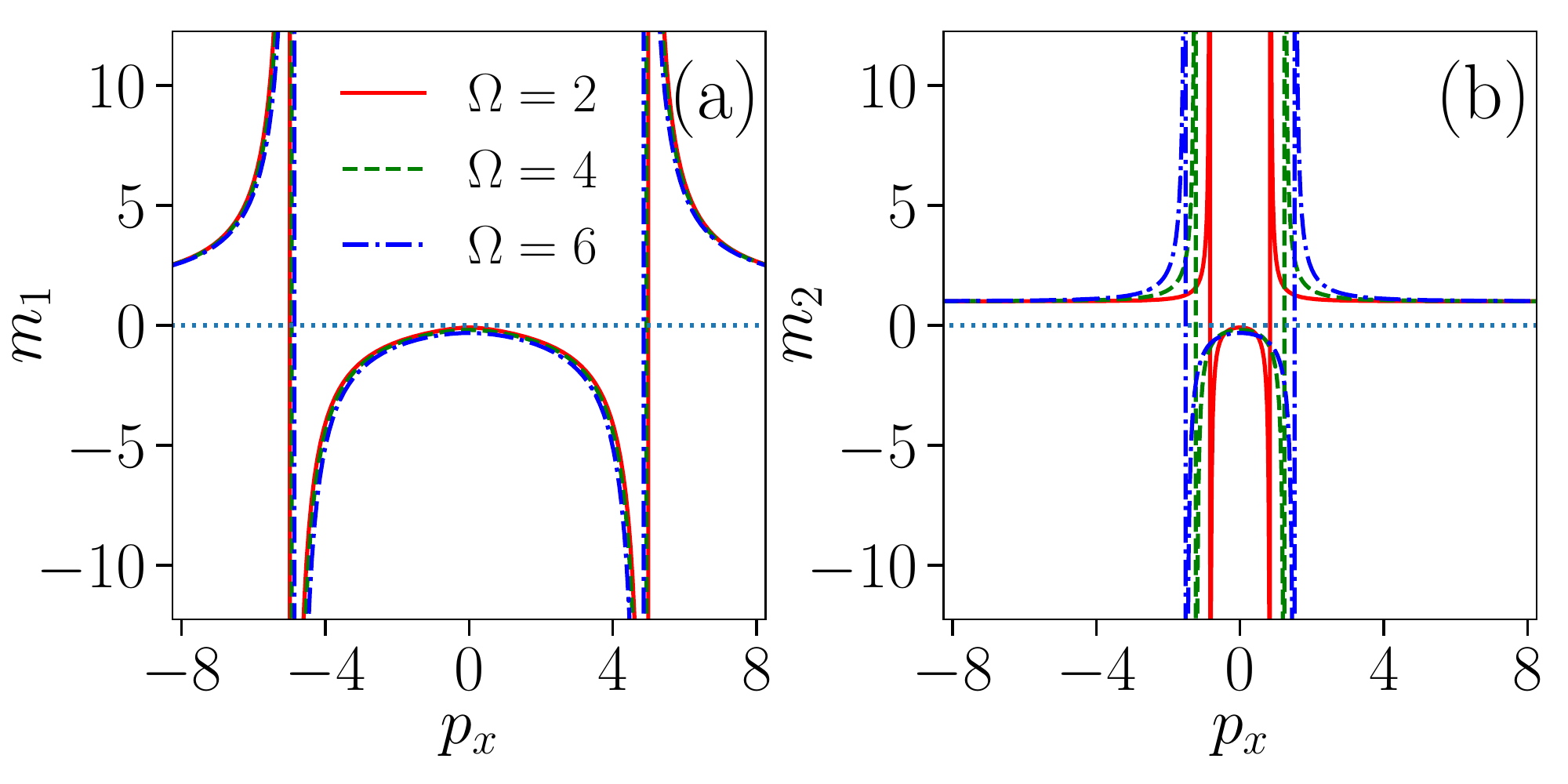}
\caption{Plots of (a) the variation of mass parameter $m_1$ and (b) the variation of mass parameter $m_2$ as a function of momentum are shown for a selected set of Rabi coupling strengths with fixed parameters $m=1$ and $k_L=5$.}
\label{m1m2}
\end{figure}

\begin{figure}[!ht]
\begin{center}
\includegraphics[width=0.65\linewidth]{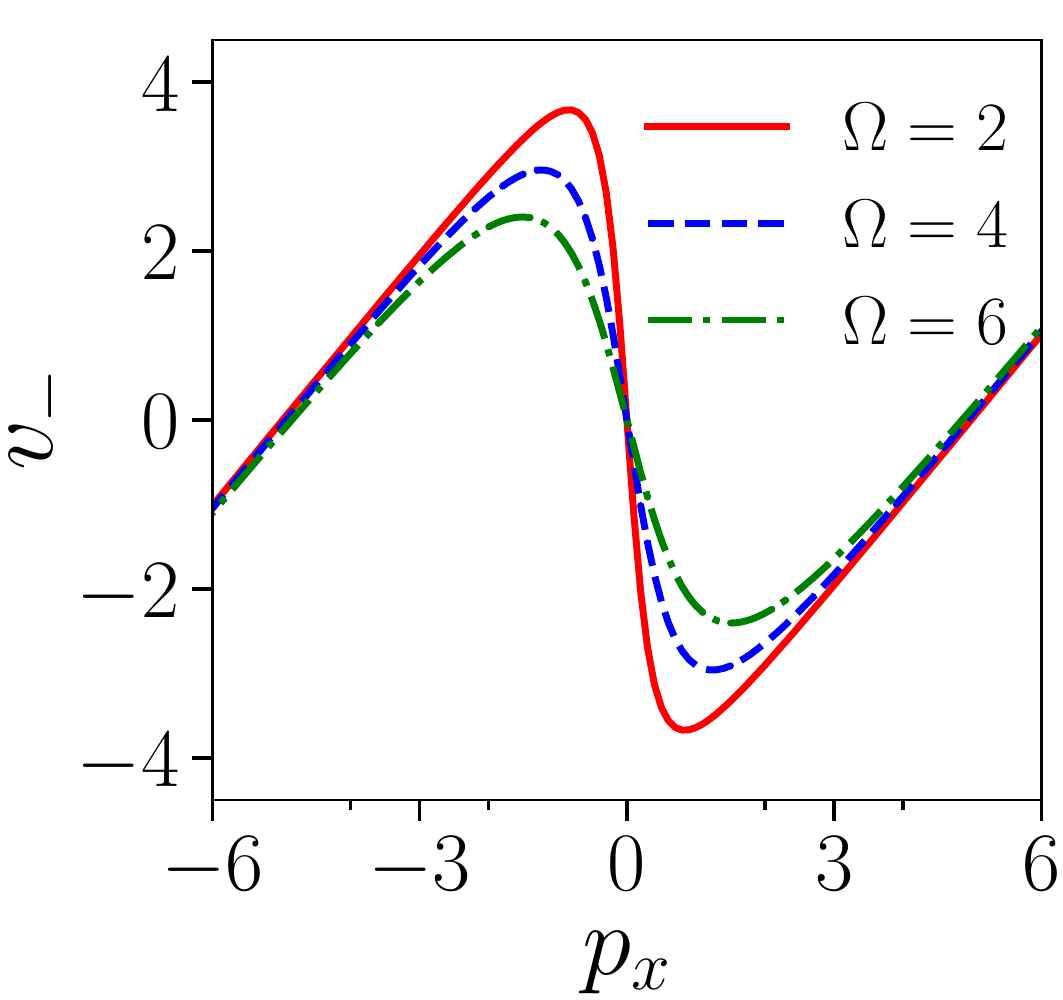}
\end{center}
\caption{Plot of group velocity as a function of $p_x$ with fixed spin-orbit coupling $(k_L = 5)$ and mass $(m=1)$ for different Rabi coupling strengths $(\Omega)$.}
\label{groupvelocityfig}
\end{figure}

\begin{figure}[!ht]
\centering\includegraphics[width=0.99\linewidth]{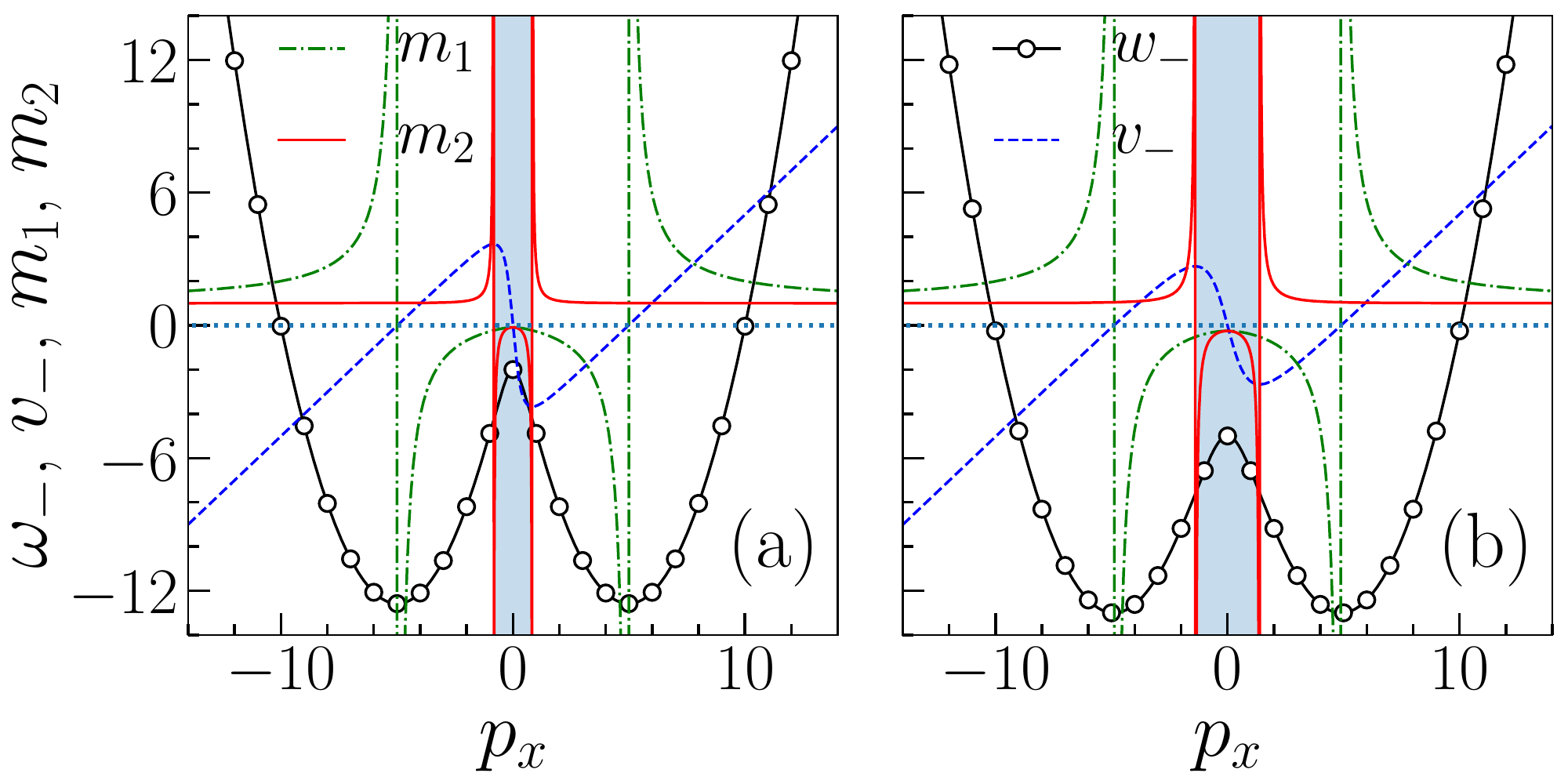}
\caption{Plots showing the variation of energy (solid black line with circles), group velocity (dashed blue line), and mass parameters ($m_1$ -- dash-dotted green line and $m_2$ -- solid red line) in momentum space  presented for two different scenarios: (a) $k_L = 6$ and $\Omega=2$, and (b) $k_L = 5$ and $\Omega=5$, with a fixed parameter of $m=1$.}
\label{Spectrum}
\end{figure}

From the general expression given by Eq.~\eqref{massparameters}, one can compute the mass parameters corresponding to the underlying system as %
\begin{subequations}
\begin{align}
m_{1} & = \frac{m}{\left(1-\frac{m k_{L}^{2}}{\sqrt{\Omega^{2}+p_x^2 k_{L}^{2}}}\right)}, \\
m_{2} & = \frac{m}{\left(1-\frac{k_{L}^{2}\Omega ^{2}m\sqrt{\Omega ^2 +p_x^2 k_{L}^{2}}}{\left(\Omega^{2}+p_x^{2}k_{L}^{2}\right)^2}\right)}.
\end{align}
\end{subequations}%
As the dispersion regime and other related characteristics critically depend on the strength of the Rabi coupling, we show, in Figs.~\ref{m1m2}(a) and \ref{m1m2}(b), the effective mass parameters in the momentum space for some representative values of Rabi coupling. %
It is evident from Fig.~\ref{m1m2} that the effective mass is sensitive to the Rabi strength. As the strength of Rabi coupling increases, the negative region of both mass parameters shrinks, with $m_2$ being the most sensitive to $\Omega$ than $m_1$. The variation of the group velocity as a function of the momentum is portrayed in Fig.~\ref{groupvelocityfig} using Eq.~\eqref{massparameters}. The absolute value of the group velocity of the wave packet is found to be
\begin{align}
v_- = \frac{p}{m}-\frac{pk_{L}^{2}}{\sqrt{\Omega^{2}+p^{2}k_{L}^{2}}}.
\label{groupvelocity}
\end{align}
It is apparent from the above discussion that the mass parameters and the group velocity of the wave packets depend on the strength of the SO and Rabi couplings which is indeed a manifestation of the potential of the SO-coupled BECs for more effective dispersion engineering on quantum systems. %
In Fig.~\ref{Spectrum}, we display the variation of the system parameters in the momentum space for some representative values of $k_L$ and $\Omega$ corresponding to the lower branch of the energy spectrum. 
The inflection point representing the change of the sign of $m_2$ can be estimated using Eq.~\eqref{m2_label} as ,
\begin{align}
{p_{x}}_{1,2}=\pm \sqrt{\left(\frac{m\Omega ^2}{k_L}\right)^{2/3}-\left(\frac{\Omega }{k_{L}}\right)^2}.
\label{infinite}
\end{align}
Similarly, the points on the momentum space at which the effective mass $m_1$ diverges can be deduced from Eq.~\eqref{m1_label} as follows
\begin{align}
{p_{x}}_{3,4}= \pm \sqrt{\frac{-\Omega^2+m^2 k^4_L}{k_L}}.
\label{divergence}
\end{align}
In the parametric space of interest, one of the notable characteristic differences between the spin-$1/2$ reported in Ref.~\cite{Colas2018} and the present case of $F=1$ is the shape of the energy spectrum. The energy dispersion spectrum is symmetric in the momentum space, as is evident from Fig.~\ref{Spectrum} unlike in Ref.~\cite{Colas2018}. It should be emphasized that the shape of the energy spectrum dictates the dynamics of the system under investigation. One can understand from Fig.~\ref{Spectrum} drawn out for two different sets of parameters that the energy spectrum, group velocity, and mass parameters depend on the strengths of Rabi coupling and SOC which means the shape of the energy spectrum can be manipulated accordingly leading to dispersion engineering.

It is worth observing from the above that one can interpret most of the dynamical behaviors of the system from the momentum-dependent velocity $v(p_x)$ given by Eq.~\eqref{groupvelocity},  which can be straightforwardly related to the mass parameters $m_1$ and $m_2$ as defined by the expression \eqref{massparameters}. Fig.~\ref{Spectrum} shows how the group velocity, energy, and mass parameters vary in momentum space as a function of Rabi coupling and SOC.  It can be observed from Figs.~\ref{Spectrum}(a) and \ref{Spectrum}(b) that the linear part on either side of the zero momentum when the mass parameter $m_2$ diverges corresponds to a local maximum or minimum in the group velocity. The maximum and minimum are sensitive to the strength of the Rabi coupling as predicted earlier from Fig.~\ref{groupvelocityfig}. In addition,  the point where the velocity becomes  zero corresponds to the maxima  of the energy dispersion curve, and the region where the velocity attains negative values corresponds to the domain where the energy dispersion curve tends to its minima. One of the key features of the velocity curve is the existence of the negative region owing to the negative mass parameters where the wavepacket moves in the opposite direction in response to the impulse.

\section{Numerical simulations}
\label{sec4}
We begin our simulation with an expanding spin-orbit coupled three-component spinor condensate with the time integration of the equation using the split-step method~\cite{Muruganandam2009, Muruganandam2021}. The negative mass regime which has been associated with several interesting phenomena like  self-interference pattern (SIP), self-trapping, etc., had earlier been  observed by David Colas et al. for spin-$1/2$ spin-orbit coupled Bose-Einstein condensates by initially positioning the condensate at the bottom of the lower branch and releasing the trap from one side which is given by single band Gross-Pitaevskii equation~\cite{Colas2018, Khamehchi2017}. 

We adopt the same approach  for spin-orbit coupled spinor $F=1$  condensates and observe some of the dynamical phenomena like solitons , dynamical instability, pileup, self-interference pattern and self-trapping effects when the condensate starts to expand in the three different component densities ($\lvert \psi_{+1} \rvert^2$, $\lvert \psi_{0} \rvert^2$ and $\lvert \psi_{-1} \rvert^2$).  To start with, we prepare the ground state wavefunctions of 1D harmonically trapped condensate by taking the initial condition as a Gaussian wave employing  the  imaginary-time propagation of the Gross-Pitaevskii equation (\ref{eq:gpe}). We study the dynamics by initially positioning the condensate at the bottom of the lower branch and releasing it from the harmonic trap during real-time propagation~\cite{Qu2017}.

\subsection{Expansion dynamics and self-interference}
In a recent development, Su et al. demonstrated the self-interfering dynamics of a wave packet using the Wigner distribution function~\cite{Su2022}. They observed the self-interfering dynamics in a non-interacting condensate by engineering dispersion using either optical lattice or spin-orbit coupling. Additionally, they observed asymmetric expansion dynamics by positioning the wave packet at the center ($x=0$) keeping detuning nonzero while symmetric expansion is witnessed for zero detuning.

 Here, density fluctuations are observed from the center to the tail of the condensate. This accumulation of density fluctuations, often referred to as ``dynamical instability'' at the edges, is called ``pileup'' and it occurs for different Rabi coupling strengths. The density fluctuations are much more pronounced when we increase $\Omega$ from $2$ to $6$ while keeping $k_L=5$.

Moreover, the expansion of the condensate depends on the strength of the nonlinearities. For smaller nonlinearities, the wave packets expand faster and continuously. However, for larger nonlinearities, by increasing the number of atoms, the expansion slows down, and the condensate stops expanding and becomes self-trapped, similar to the case of Bose-Einstein condensates in optical lattices by Wang et al., who studied larger nonlinearities with different atom numbers \cite{Wang2006}.

\begin{figure}[!ht]
\centering\includegraphics[width=0.99\linewidth]{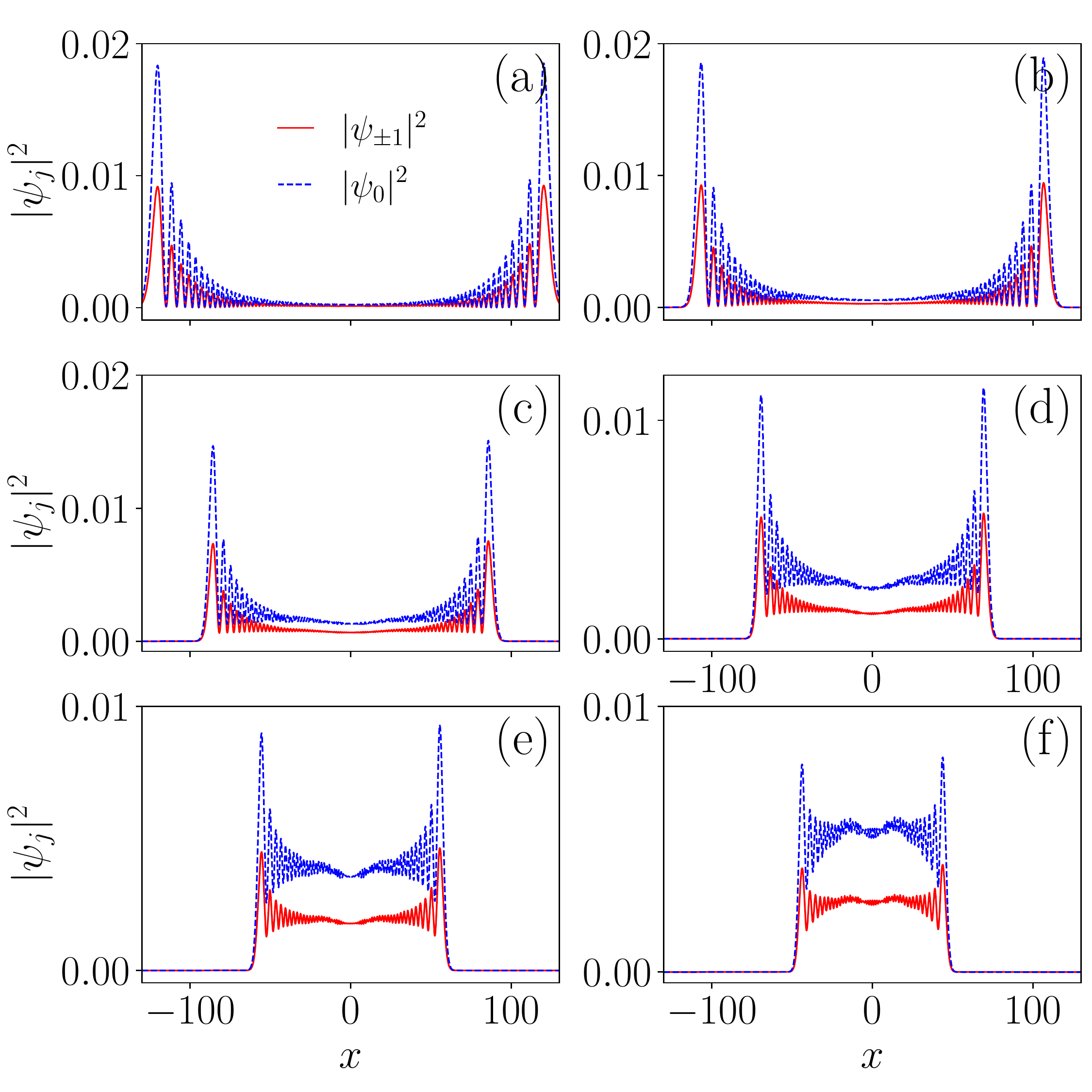}
\caption{The snapshots of the one-dimensional symmetric expansion of a spin-orbit coupled Bose-Einstein condensate without magnetization ($M=0$) for $k_L=5$ and for different strengths of (a) $\Omega=1$, (b) $\Omega=2$, (c) $\Omega=4$,  (d) $\Omega=6$, (e) $\Omega=8$, and (f) $\Omega=10$ at $t=30$.}
\label{fig:OM}
\end{figure}

We first study self-interference packet for the repulsive condensates with nonlinearities $c_0 = 0.25$ and $c_2 = -0.001$. Fig.~\ref{fig:OM} shows the condensate expansion for different values of $\Omega$ while keeping $k_L=5$ and magnetization at zero. %
We observe a symmetric expansion where the densities of $\lvert \psi_{+1} \rvert^2$ and $\lvert \psi_{-1} \rvert^2$ are equal, while $\lvert \psi_{0} \rvert^2$ has a different density. The snapshots at $t=30$ are depicted in Figs. \ref{fig:OM}(a)-(f). Fig. \ref{fig:OM}(a) shows fluctuations only at the edges when $\Omega=1$. In contrast, Fig. \ref{fig:OM}(b) exhibits a decrease in density while the fluctuations begin to appear from the center to the edges, and the expansion slows down compared to Fig. \ref{fig:OM}(a). The density is much smaller for $\Omega=6$, $\Omega=8$, and $\Omega=10$, as shown in Figs. \ref{fig:OM}(d) - \ref{fig:OM}(e), and the fluctuations are observed from the center to the edges. Here, the expansion is slower compared to the other cases. The slower expansion rate of the condensate leads to an increase in the effective mass, which is identical to that of Ref. \cite{Qu2017}.

It can also be observed from Fig.~\ref{fig:OM} that when we release the trap and allow the condensates (wave packet) to expand by increasing the strengths of Rabi coupling, the width and density of the wave packet decrease. This ultimately leads to the trapping of the atoms within the condensates, a phenomenon which we refer to as ``self-trapping.'' The word ``self-trapping'' arises because we switch off the trap which allows the condensates to trap the atoms inside by increasing the Rabi coupling. The self-trapping phenomenon in BECs has many applications, such as matter-wave interferometry, atom optics, and quantum information processing~\cite{Andrey2010}. It is worth pointing out at this juncture that the introduction of magnetization does not impact the dynamics of self-trapped BECs as well.

\begin{figure}[!ht]
\centering\includegraphics[width=0.99\linewidth]{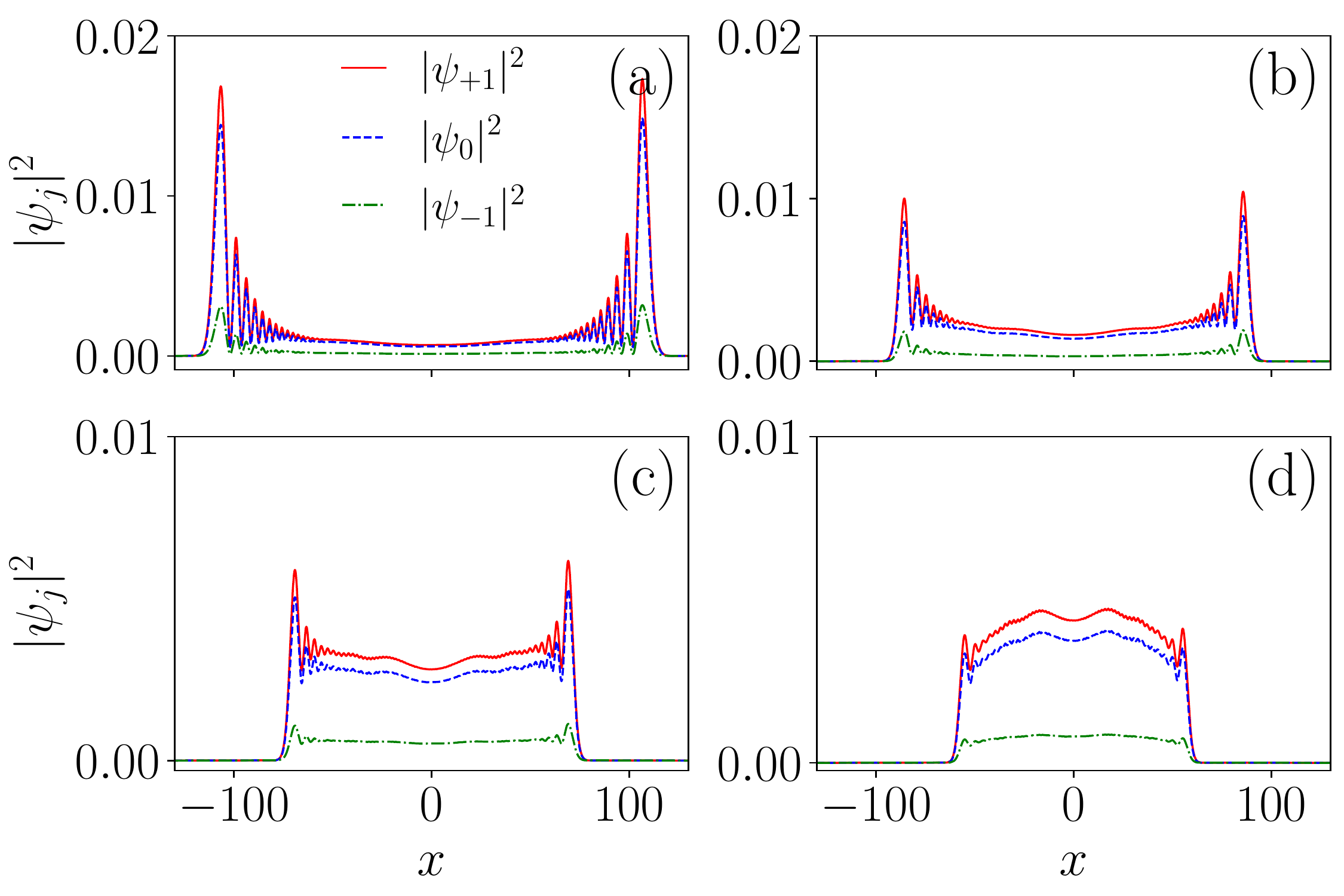}
\caption{The snapshots  of the one-dimensional symmetric expansion of a spin-orbit coupled Bose-Einstein condensate with magnetization ($M=0.4$) for $k_L=5$ and for different strengths of (a) $\Omega=2$, (b) $\Omega=4$, (c) $\Omega=6$, (d) $\Omega=8$ at $t=30$.}
\label{fig:M-OM}
\end{figure}

Similarly, Fig.~\ref{fig:M-OM} shows the condensate expansion for different strengths of $\Omega$, with $k_L=5$ and a fixed magnetization of $M = 0.4$. %
All these figures are similar to Fig.~\ref{fig:OM}, except that three distinct densities are observed for each case due to the removal of degeneracy, while the other observations remain the same as before.

Using real-time propagation, we perform a numerical simulation of the expansion of a quasi-1D ferromagnetic Bose-Einstein condensate (BEC) by releasing it from a harmonic trap. %
\begin{figure}[!ht]
\centering
\includegraphics[width=\linewidth]{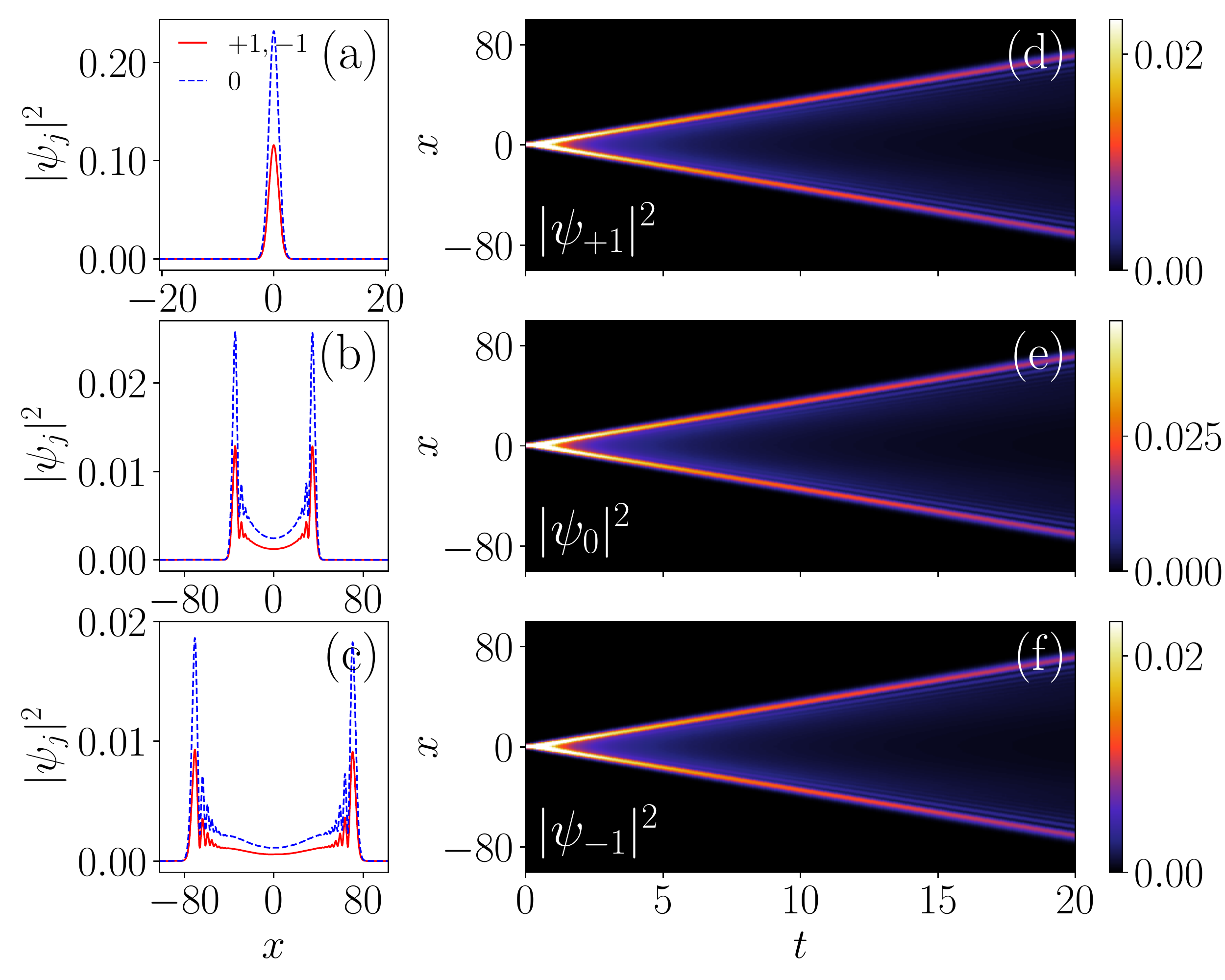}
\caption{Plots showing the expansion dynamics with coupling parameters $\Omega = 2$ and $k_L = 5$, numerically calculated with nonlinearities $c_0 = 0.25$ and $c_2 = -0.001$, for the case without magnetization. The 1D density plot at times $t = 0$, $10$, and $20$ is shown in panels (a), (b), and (c), respectively. Panels (d), (e), and (f) show the dynamics of the spin component densities $\lvert \psi_{+1} \rvert^2$, $\lvert \psi_{0} \rvert^2$, and $\lvert \psi_{-1} \rvert^2$.}
\label{fig:num1}
\end{figure}%
The densities of the spin components, $\lvert \psi_{j} \rvert^2$ ($j=0, \pm{1}$), at three different time instances are depicted in Figs.~\ref{fig:num1}(a) - \ref{fig:num1}(c). Figs.~\ref{fig:num1}(d) - \ref{fig:num1}(f) present a space-time plot of the densities, showing that the self-interaction causes the momentum distribution to broaden, making them spread in real space. In the absence of magnetization, $\psi_{+1}$ and $\psi_{-1}$ have equal pseudo spin component densities. However, the presence of magnetic field breaks the spin-1 component's degeneracy, leading to lower density fluctuations at the edges~\cite{Anker2005}. Figs.~\ref{fig:num1}(d) and \ref{fig:num1}(f) exhibit the same density for $\psi_{+1}$ and $\psi_{-1}$, whereas Fig.~\ref{fig:num1}(e) shows a different density, $\psi_{0}$, when the magnetization is zero.

\begin{figure}[!ht]
\centering\includegraphics[width=\linewidth]{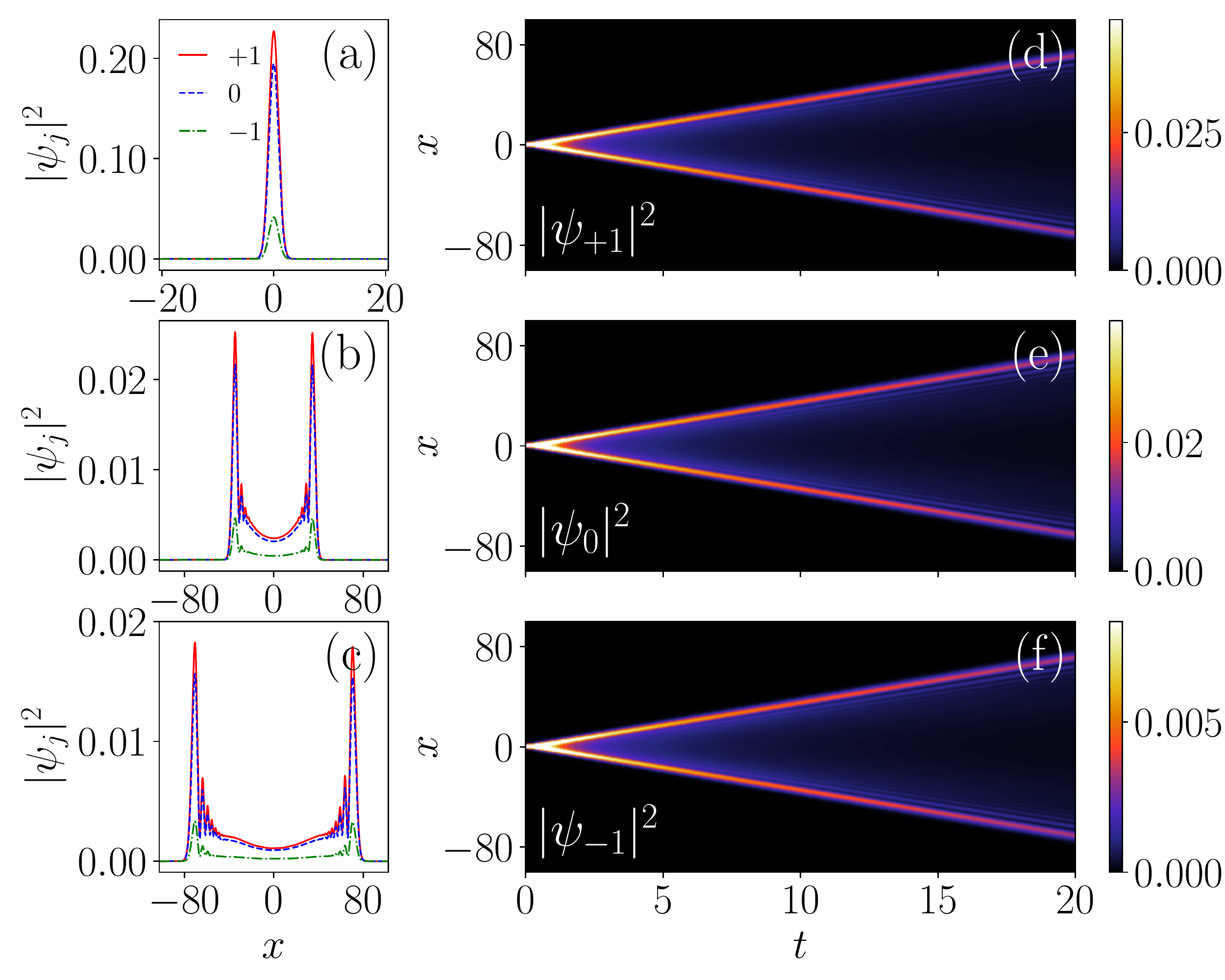}    
\caption{Plots displaying the real-time expansion of a Bose-Einstein condensate (BEC) with coupling parameters $\Omega = 2$ and $k_L = 5$, as computed numerically with nonlinearities $c_0 = 0.25$ and $c_2 = -0.001$, and magnetization of $0.4$ at different times: (a) $t = 0$, (b) $t = 10$, and (c) $t = 20$. The space-time plot shows the dynamics of the densities for (d) $\lvert \psi_{+1} \rvert^2$, (e) $\lvert \psi_{0} \rvert^2$, and (f) $\lvert \psi_{-1} \rvert^2$.}
\label{fig:num2}
\end{figure}

Similarly, SIP is also observed in Figs.~\ref{fig:num2}(a) to \ref{fig:num2}(c), which depict the component densities $\lvert \psi_{j} \rvert^2$ ($j = 0$, $\pm{1}$) at three different instants of time. The space-time plot of the densities is shown in Figs.~\ref{fig:num2}(d) to \ref{fig:num2}(f) when the magnetization is $0.4$. 

\begin{figure}[!ht]
\centering\includegraphics[width=\linewidth]{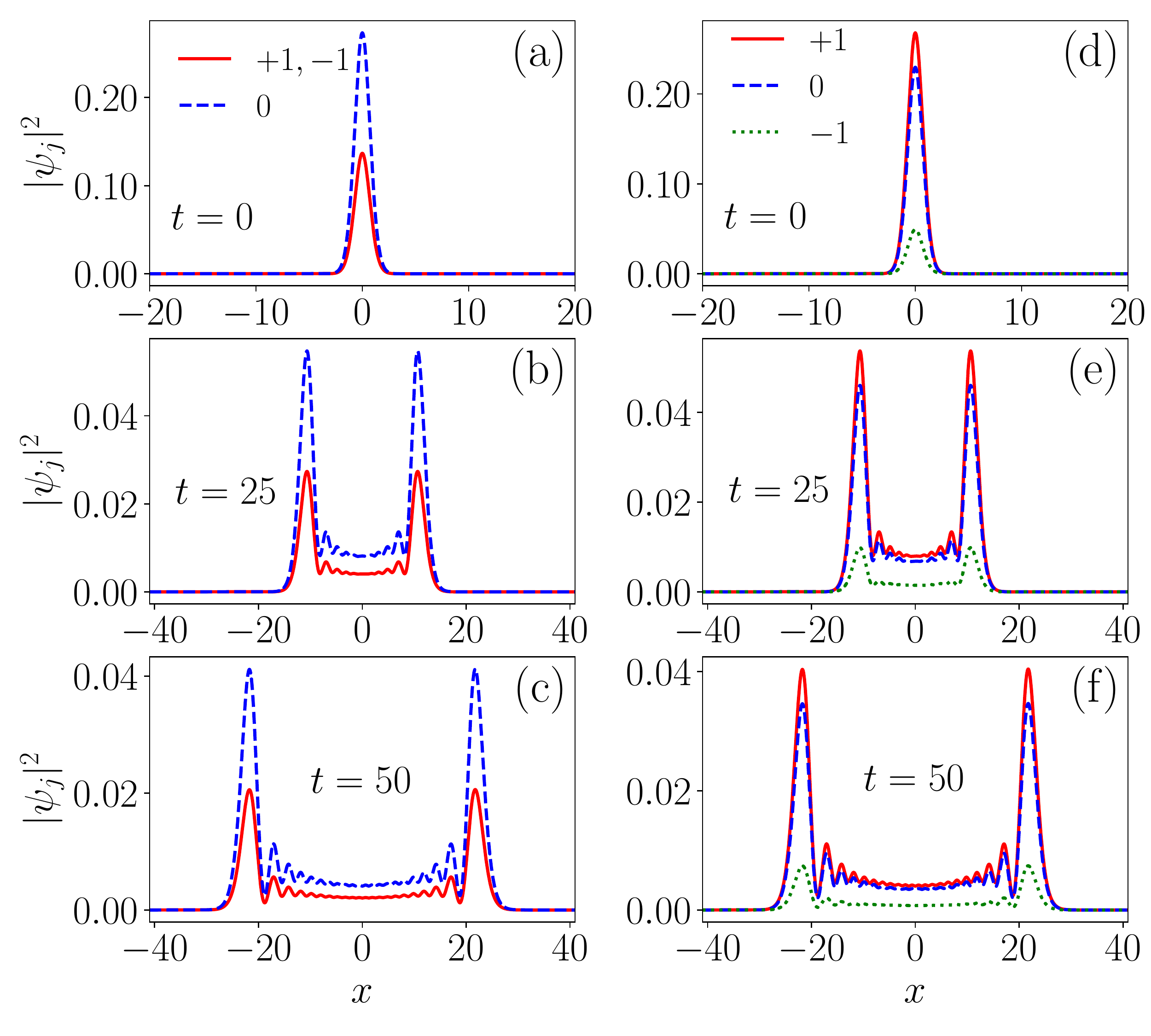}
\caption{Plots illustrating the evolution of spin-component densities, $\lvert\psi_j\rvert^2$, for coupling parameters $\Omega=2$ and $k_L=5$ and interaction strengths of $c_0=-1.5$ and $c_2=-0.3$ at different instances of time: (a)-(c) depict the scenario where magnetization is absent, while (d)-(f) corresponds to the case where magnetization is present. The time intervals shown are $t=0$, $t=25$, and $t=50$.}
\label{fig:num3}
\end{figure}

We have also observed a SIP for attractive interactions in Fig.~\ref{fig:num3}, which shows the numerically calculated real-time propagation of the three component densities $\psi_{j}$ where $j = 0, \pm 1$ of a quasi-1D ferromagnetic BEC with the release of the harmonic trap, with nonlinearities being $c_0 = -1.5$ and $c_2 = -0.3$.
Figs.~\ref{fig:num3}(a), \ref{fig:num3}(b), and \ref{fig:num3}(c) show the expansion dynamics when the magnetization is zero, which emphasizes that the density expansion is the same for $\psi_{+1}$ and $\psi_{-1}$, whereas it is different for $\psi_{0}$. 
On the other hand, Figs.~\ref{fig:num3}(d), \ref{fig:num3}(e), and \ref{fig:num3}(f) correspond to the expansion dynamics when the magnetization is $0.4$, demonstrating that the density expansion is different for all three cases, which indicates the removal of degeneracy. Moreover, one can also observe a clear expansion of the self-interference packet with time evolution.
\begin{figure}[!ht]
\centering\includegraphics[width=\linewidth]{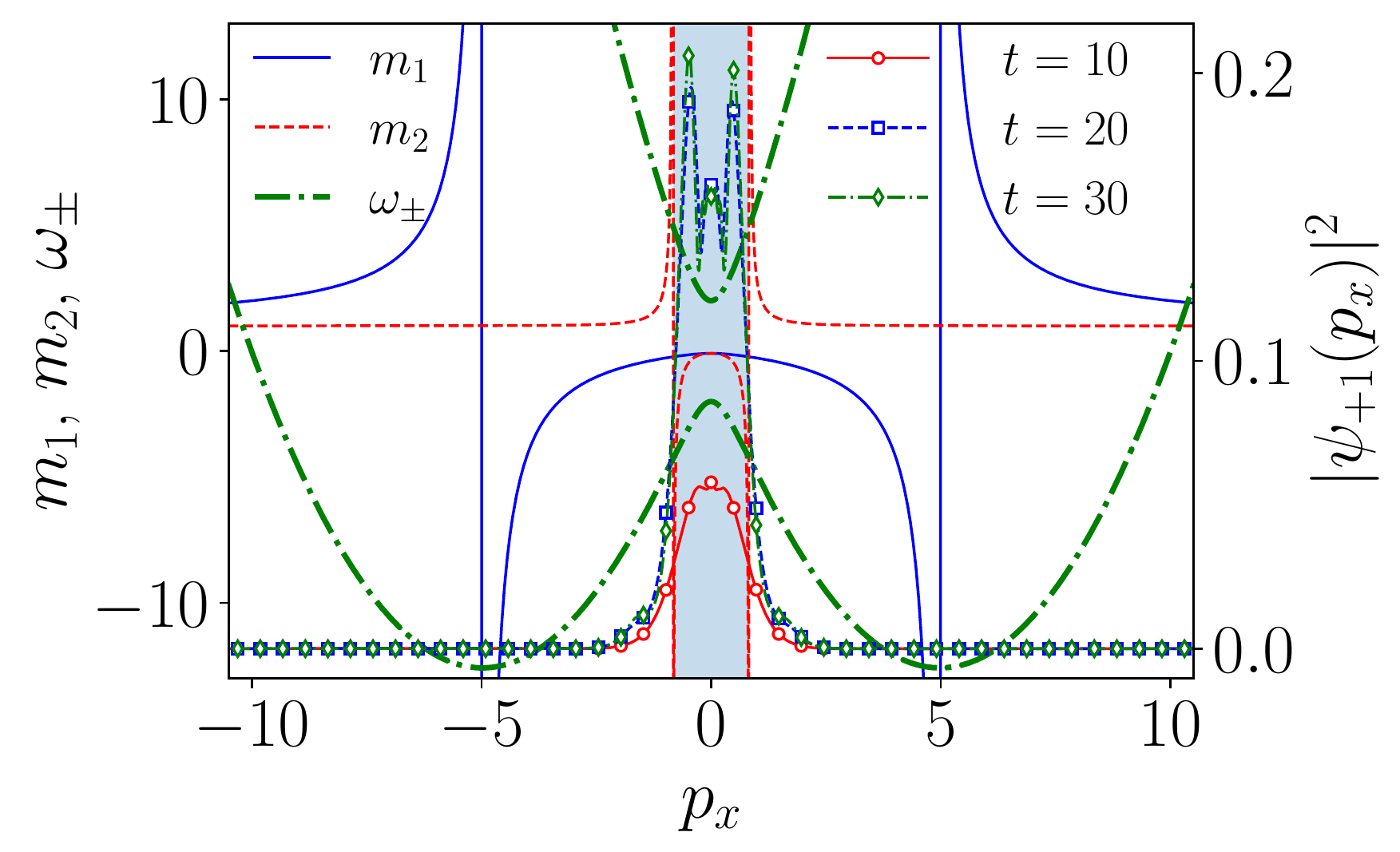}
\caption{Plot of the density of the expanding wave packet in the momentum space, $\psi_{+1}(p_x)$ at $t = 10$, $20$ and $30$. The parameters are fixed at $c_0 = 0.25$ and $c_2 = -0.001$, $\Omega = 2$ and $k_L = 5$ [cf. Fig.~\ref{fig:num1}].}
\label{fig:super}
\end{figure}%

From the dispersion relation shown in Fig.~\ref{Parabolic}, one observes a gap between the two lowest bands around  $p_x=0$ which indicates that the system may admit gap solitons. To extract  gap solitons, one begins with the expansion dynamics shown in Fig.~\ref{fig:num1} and transform it into the momentum space. We then superimpose the dispersion relation and negative masses onto the wave packet dynamics in the momentum space. From Fig.~\ref{fig:super}, one witnesses a localized wave around $p_x=0$ . This soliton is situated  at the intersection of two negative masses and is confined within a narrow interval on both sides of $p_x=0$ termed as a band gap. This gap soliton is completely different from what is being observed in \cite{Su2021} where one observes two peaks by virtue of two maxima in the triple well potential. The presence of one maxima (energy) in the dispersion relation gives rise to one peak (localized pulse) which lies at the intersection of two negative masses.

\subsection{Velocity profile during expansion and phase transition}
\begin{figure}[!ht]
\centering\includegraphics[width=\linewidth]{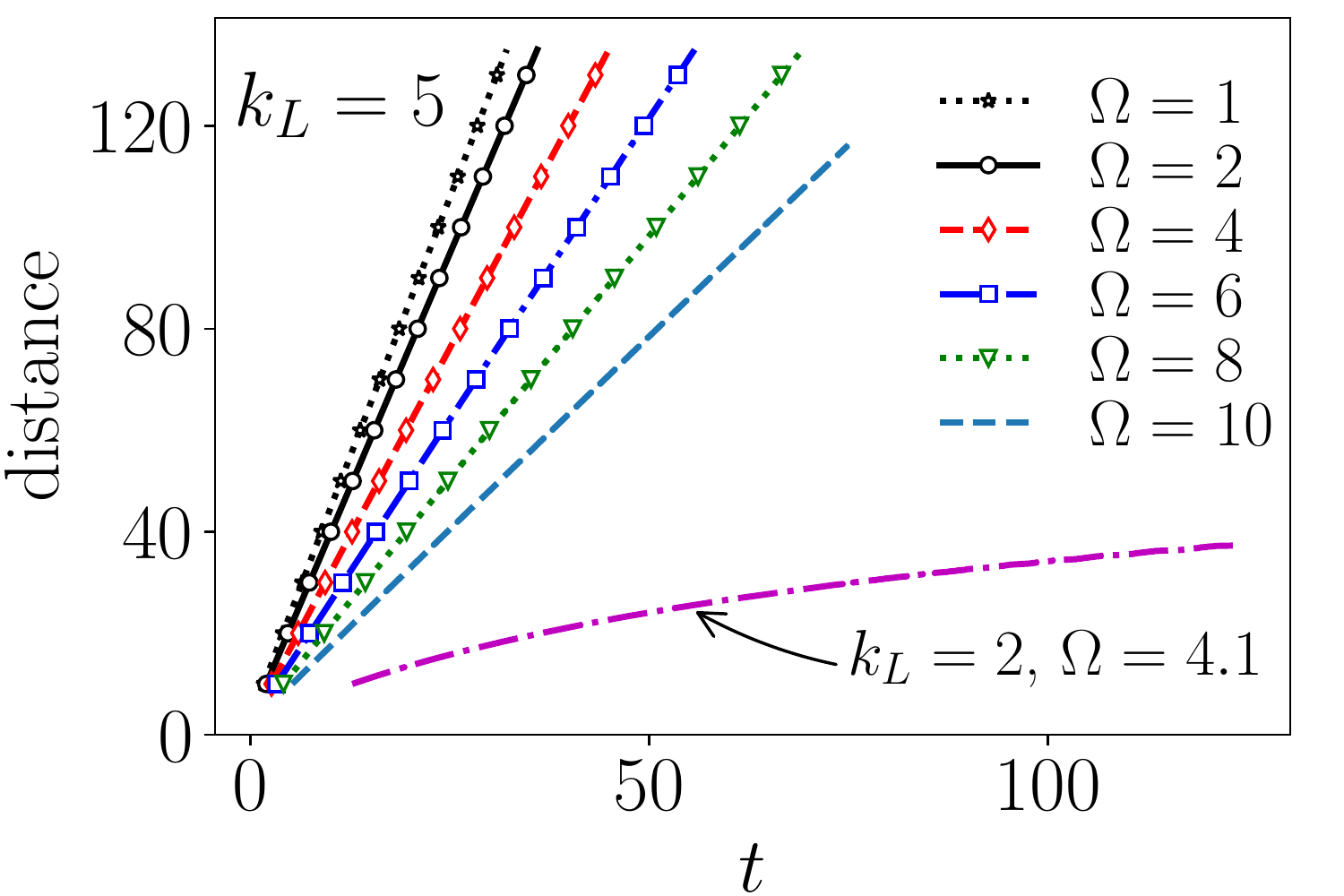}
\caption{Distance versus time plot showing the expansion for various values of $\Omega$, with $k_L$ fixed at $5$ (without magnetization). The slopes of these curves correspond to the expansion velocities and are $v = 4.10$ ($\Omega=1$), $3.64$ ($\Omega=2$), $2.94$ ($\Omega=4$), $2.38$ ($\Omega=6$), $1.92$ ($\Omega=8$) and $1.52$ ($\Omega=10$) in units $l_0 \times \omega_x$. The magenta dash-dotted line corresponds to $k_L = 2$ and $\Omega = 4.1$}.
\label{fig:velocal}
\end{figure}

Next, we shall investigate the velocity profile of the condensate during symmetric expansion. We calculate the expansion velocity during time evolution by recording the time taken for the expanding wavefront, with a threshold amplitude typically a few percent of the maximum amplitude, to cross each spatial grid point. Fig.~\ref{fig:velocal} depicts the distance versus time graph of the expanding condensate for various strengths of $\Omega$, with $k_L$ fixed at $5$ and without magnetization. 
The plot drives home the point that for small values of Rabi frequency, i.e., when $\Omega=1$ and $\Omega=2$, the wave packet expands faster in a shorter interval of time. The wave packet expansion ``slows down'' as we increase the Rabi coupling, as seen in the cases of both $\Omega=2$ and $\Omega=4$. When $\Omega= 6$, $8$, or $10$, the expansion of the condensate slows down significantly. This ``slow down'' of the wave packet is another signature of negative mass regime.  In particular, one observes a flat profile for $k_L=2$ and $\omega=4.1$. The velocity profile exhibited during the condensate expansion is comparable to the one observed in a prior study by Khamehchi et al.~\cite{Khamehchi2017}. %
\begin{figure}[!ht]
\centering\includegraphics[width=0.99\linewidth]{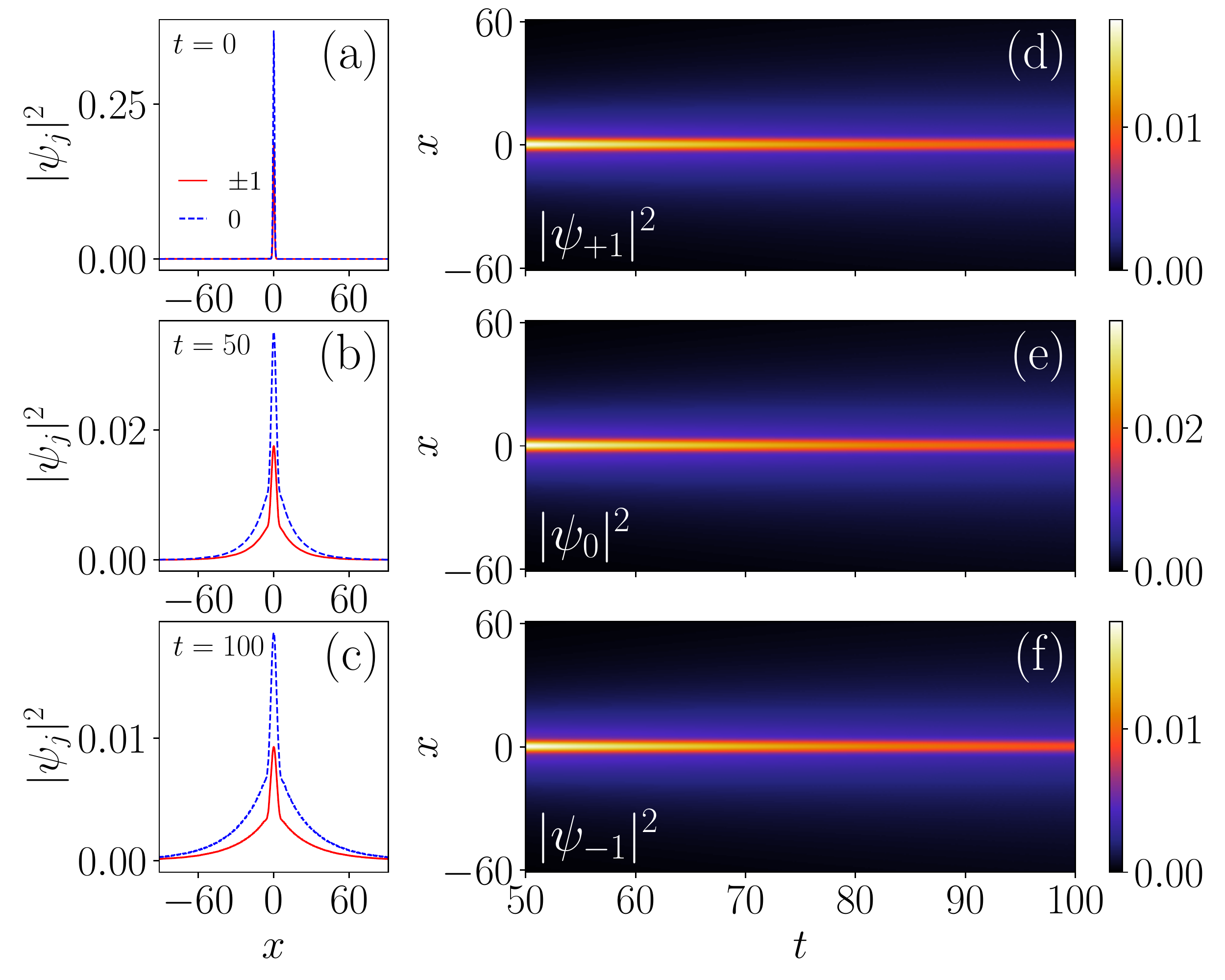}
\caption{Emergence of solitons in the  dynamics Bose-Einstein condensates (BEC) with coupling parameters $k_L = 2$ and $\Omega = 4.1$ computed numerically with nonlinearities $c_0 = 0.25$ and $c_2 = -0.001$, without magnetization at different times: (a) $t = 0$, (b) $t = 50$, and (c) $t = 100$ and the corresponding space-time plot for (d) $\lvert \psi_{+1} \rvert^2$, (e) $\lvert \psi_{0} \rvert^2$, and (f) $\lvert \psi_{-1} \rvert^2$.} 
\label{fig:selftrap}
\end{figure}%
The flat profile has motivated us to look for the time evolution of spinor BECs for the same choice of parameters. The plots depicted in Figs.~\ref{fig:selftrap}(a)-(c) showcase the dynamics of spin component densities for $k_{L}=2$ and $\Omega=4.1$, observed at various time intervals. Figs.~\ref{fig:selftrap}(d)-(f) illustrate the temporal evolution and unearth the potential emergence of a localized state resulting from the interplay of nonlinear atomic interactions with SO coupling and Rabi parameters. This phenomenon leads to the formation of a localized region with relatively high density, often referred to as a soliton. The soliton remains localized around $x=0$ with a marginal change in amplitude during the time evolution.

\begin{figure}[!ht]
\centering\includegraphics[width=0.99\linewidth]{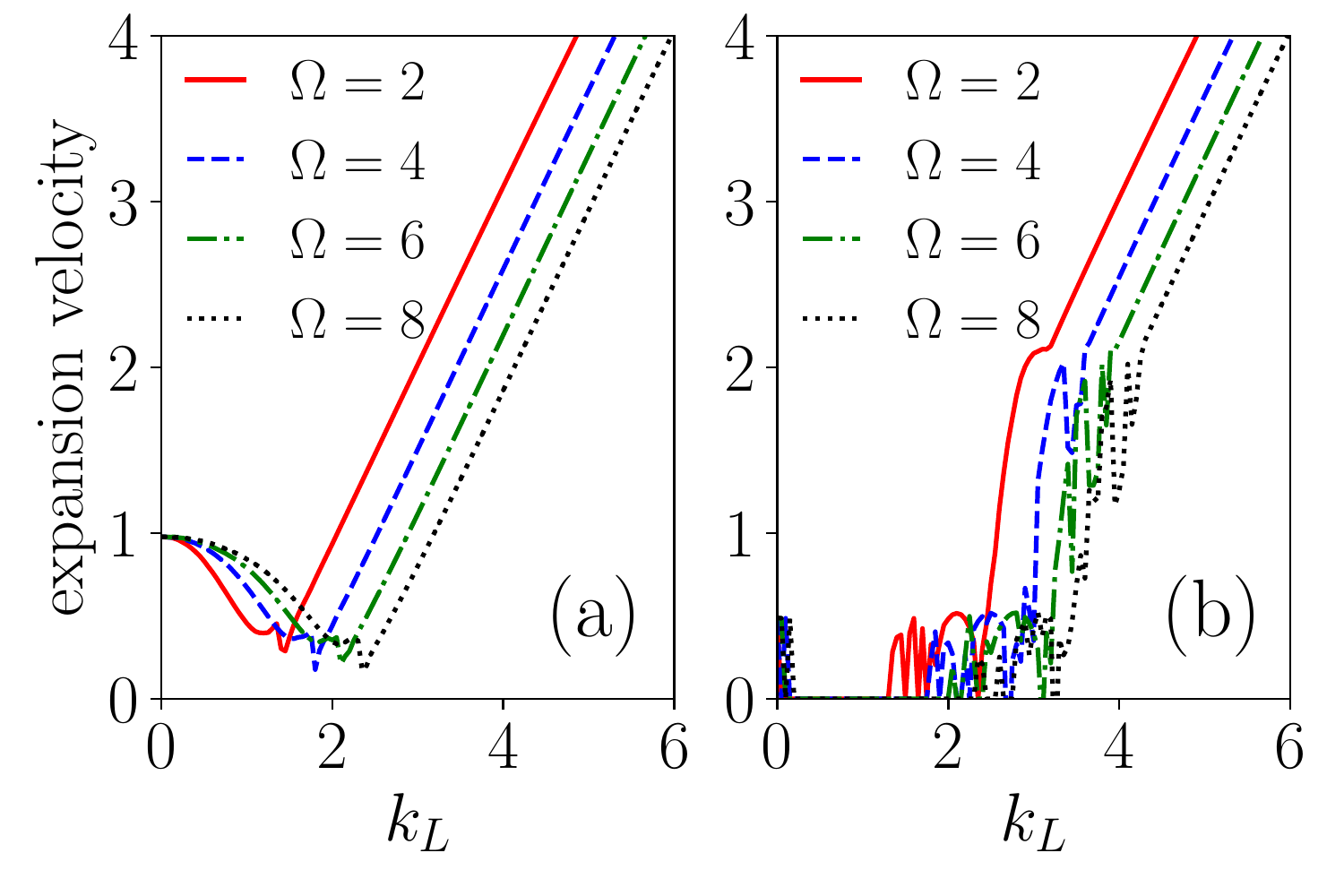}
\caption{The plot of the wave-packet expansion velocity as a function of $k_L$ for different values of $\Omega$, with (a) $c_0 = 0.25$ and $c_2 = -0.001$ (repulsive interaction) and (b) $c_0 = -1.5$ and $c_2 = -0.3$ (attractive interaction), in the absence of magnetization.} 
\label{fig:vkl}
\end{figure}

We then study the expansion velocity of the condensate for different combinations of $k_L$ and $\Omega$. Figs.~\ref{fig:vkl}(a) and \ref{fig:vkl}(b) show the variation of expansion velocity as a function of $k_L$ for different values of $\Omega$ for attractive and repulsive interactions, respectively. %
It may be observed that the velocity decreases for small values of $k_L$ and reaches a minimum close to zero at a critical value. After this critical $k_L$, the velocity increases linearly with $k_L$ which is shown in Fig.~\ref{fig:vkl}(a). %
\begin{figure*}[!ht]
\centering\includegraphics[width=0.99\linewidth]{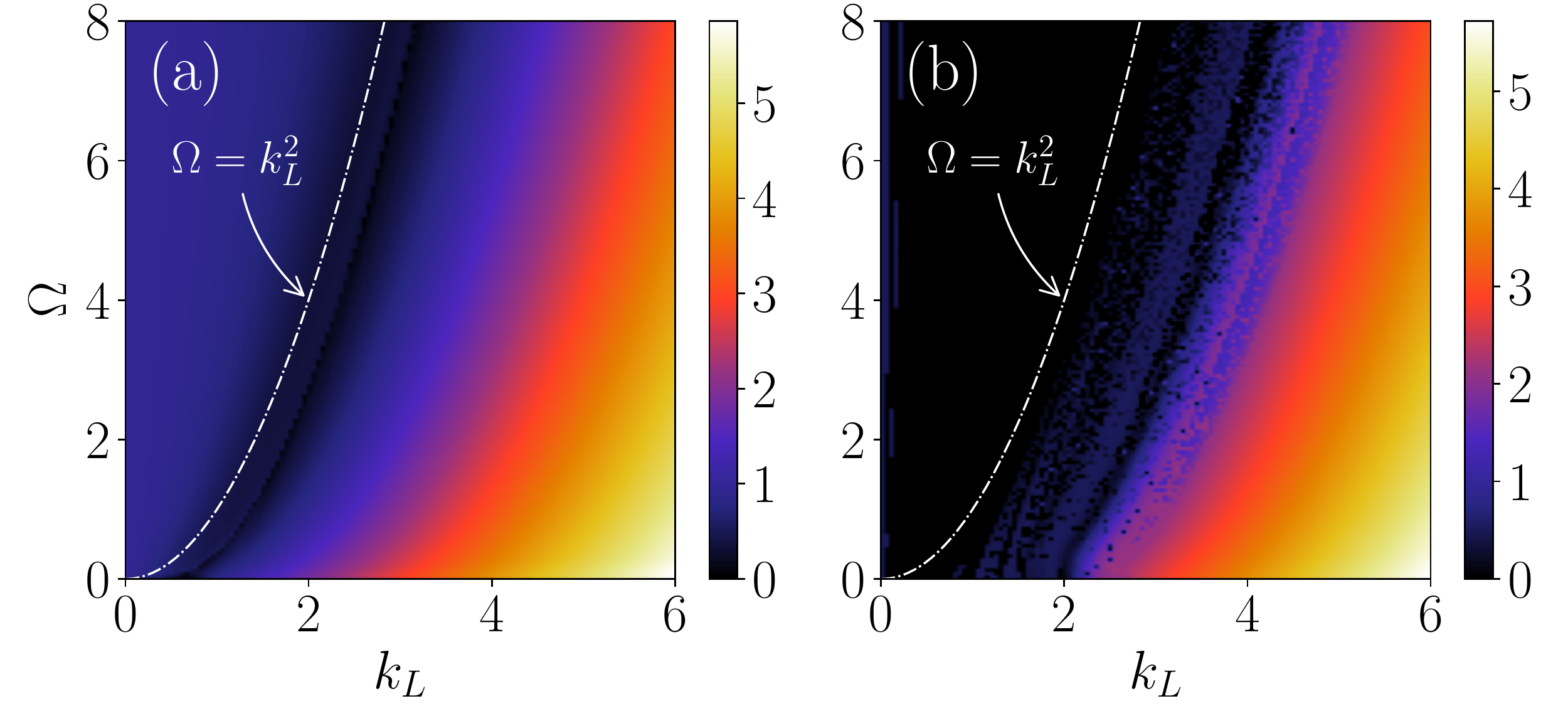}
\caption{Phase diagram illustrating the wave-packet expansion velocity as a function of $k_L$ and $\Omega$, with (a) $c_0 = 0.25$ and $c_2 = -0.001$ (repulsive interaction) and (b) $c_0 = -1.5$ and $c_2 = -0.3$ (attractive interaction), in the absence of magnetization. The theoretical $\Omega = k_L^2$ boundary, where the transition from plane wave to stripe pattern occurs, is denoted by the white-dash-dotted curve.} 
\label{fig:vpro}
\end{figure*}%
Similarly, in Fig.~\ref{fig:vkl}(b) for small values of $k_L$, the velocity is nearly equal to zero and there is a sudden increase in velocity after the critical point. Another interesting observation is that the critical value of $k_L$ that separates the condensates with different velocity profiles is almost the same except that for attractive interactions, the momentum is imparted to the condensates only after the critical value. This implies that the self-trapping of the condensates is more pronounced in attractive interactions. 

In addition, we have numerically computed the expansion velocities in the $k_L - \Omega$ plane for a range of values: $k_L \in [0, 6]$ and $\Omega \in [0, 8]$. Fig.~\ref{fig:vpro}(a) illustrates a phase diagram of the velocity profiles of the expanding wave-packet front in the $k_L - \Omega$ plane for the repulsive case with $c_{0}=0.25$ and $c_{2}=-0.001$. Based on these figures, we observe that the velocity patterns are non-uniform and vary according to the values of $k_L$ and $\Omega$.
By varying $k_L$ and $\Omega$, we have observed that the expansion velocity approaches zero for certain critical values of these parameters. At the critical point, there is an effective localization. This observation suggests the occurrence of a quantum phase transition, which is actually from the plane wave to the stripe wave phase. To determine the critical values of $k_L$ and $\Omega$, we have examined the density profiles of the condensate. The values we obtained are consistent with those computed using the single-particle dispersion relation described in Eq.~\eqref{energy-dis}. The dispersion relation predicts a plane wave (PW) phase for $\Omega < k_L^2$ and a stripe wave (SW) phase for $\Omega > k_L^{2}$.

Similarly, in Fig.~\ref{fig:vpro}(b), we show the velocity profiles of the expanding wave-packet front as a function of the strengths of the SO and Rabi couplings, $k_L$ and $\Omega$, for the attractive case with $c_{0}=-1.5$ and $c_{2}=-0.3$. We can see that the velocity is equal to zero, and the condensate does not expand for small values of the SO coupling strengths. However, for large $k_L$, the velocity patterns resemble those of the repulsive case shown in Fig.~\ref{fig:vpro}(a). Since the system is attractive, it is quite natural to observe self-trapping for small values of $k_L$. The phase transition from the plane wave phase to the stripe phase occurs close to the analytical prediction, which is $\Omega=k_{L}^2$. Similar phase transitions have been reported for harmonic traps in previous works~\cite{Zhang2013, Ravisankar2020, Cabedo2021}.

\section{Conclusion} 
\label{sec5}
In this paper, we have investigated the dynamics of $F=1$ spinor spin-orbit coupled BECs described by a three-coupled GP equation in a range where the effective mass becomes negative due to the interplay between SO coupling and Rabi coupling. The density and time evolution profiles of the condensates show the existence of self-interfering packets (SIPs) for different coupling parameters for magnetized and unmagnetized states for both repulsive and attractive interactions. The density fluctuations are observed from the center to the tail, and we have found that increasing $\Omega$ while keeping $k_{L}$ constant reduces the expansion. Our investigation highlights the observation of a symmetric double-well potential that identifies two stable regimes, in contrast, to spin-$1/2$ SOC BECs. Our results suggest the possibility of observing multiple stable states based on the SO coupling, Rabi coupling, and interaction strength. Furthermore, we have observed a quantum phase transition from a plane wave phase to a stripe wave phase, where the expansion velocity becomes nearly zero. This transition may have wider ramifications in the areas of quantum and condensed matter physics. Besides, several signatures associated with the negative mass regime, such as symmetric expansion, pile-up, modulation instability, slow down,  self-trapping and gap solitons have also been identified, in addition to SIP. We believe that the results of our paper may motivate researchers to investigate the implications of negative masses in spinor $F=1$ spin-orbit coupled BECs from an experimental perspective. The results of the above investigation may motivate researchers to explore challenging concepts of the cosmos, such as wormholes, cosmic voids, time travel to the past etc., in the near future.

\acknowledgments
KR acknowledges financial support from UGC-SJSGC. SB and RR wish to thank the Council of Scientific and Industrial Research (CSIR), the Government of India for the financial support under Grant No. 03(1456)/19/EMR-II. The work of P.M. is supported by DST-SERB under Grant No. CRG/2019/004059, FIST (Department of Physics), and MoE RUSA 2.0 (Physical Sciences) Programmes.

%
\end{document}